\documentstyle[eqsecnum,preprint,aps,epsfig,floats]{revtex}
\begin{document}
\hfill{NCKU-HEP-99-01}\par
\hfill{IRB-TH-1/99}\par
\hfill{hep-ph/9902205}\par
\vskip 1.0cm
\centerline{\large\bf Determination of heavy-meson wave functions}\par
\centerline{\large\bf from $B$ decays}
\vskip 0.5cm
\centerline{Hsiang-nan Li$^a$ and Bla\v zenka Meli\'c$^b$}
\vskip 0.3cm
\centerline{$^a$ Department of Physics, National Cheng-Kung University,}
\centerline{Tainan, Taiwan 701, Republic of China}
\vskip 0.3cm
\centerline{$^b$ Theoretical Physics Division, Rudjer Bo\v skovi\'c
Institute}
\centerline{P.O. Box 1016, HR-10001 Zagreb, Croatia}

\begin{abstract}

We extract the $B$, $D$, and $D^*$ meson wave functions from the CLEO data
of the decays $B\to K^*\gamma$ and $B\to D^{(*)}\pi$ in the perturbative
QCD framework. In this formalism, various logarithmic corrections are
organized to give the Wilson evolution from the $W$ boson mass down to the
characteristic scale of a decay process, which is of order of the $b$ 
quark mass, and the Sudakov evolution from the characteristic scale to a 
lower factorization scale of order $\Lambda_{\rm QCD}$. With large
logarithms organized, the $b$ quark decay amplitudes are evaluated reliably
in perturbation theory. Below the factorization scale, QCD dynamics is
regarded as being nonperturbative, and absorbed into meson wave functions.
Because of their universality, the heavy-meson wave functions determined in 
this work, can be employed to make predictions of other decay modes. We also 
observe that the dependence of heavy meson wave functions on intrinsic 
parton transverse momenta plays an important role in the explanation of data.

\end{abstract}


\newpage

\section{Introduction}

Recently, a PQCD factorization theorem for exclusive heavy-meson decays has 
been developed by one of us \cite{CL}, and applied to the nonleptonic
$B\to D^{(*)}\pi(\rho)$ decays \cite{YL} and the penguin-induced radiative
$B\to K^*\gamma$ decays \cite{LL}. These
decays involve three scales: the $W$ boson mass $M_W$, at which the
matching conditions of the effective weak Hamiltonian to the full
Hamiltonian are defined, the typical scale $t$ of a hard amplitude,
which reflects the specific dynamics of a heavy-meson decay, and the
factorization scale $1/b$, with $b$ being the conjugate variable of parton
transverse momenta. The dynamics below $1/b$ is regarded as being
completely nonperturbative, and parametrized into meson wave functions
$\phi(x,b)$, $x$ being the momentum fraction. Above the factorization scale,
PQCD is reliable and radiative corrections produce two types of large
logarithms: $\ln(M_W/t)$ and $\ln(tb)$. The former are summed by
renormalization-group (RG) equations to give the evolution from $M_W$ down
to $t$ described by the Wilson coefficients $c(t)$, while the latter are
summed to give the evolution from $t$ to $1/b$.

There also exist double logarithms $\ln^2(Pb)$ from the overlap of
collinear and soft divergences in radiative corrections to meson
wave functions, $P$ being the dominant light-cone component
of a meson momentum. The resummation of these double logarithms leads to a
Sudakov form factor $\exp[-s(P,b)]$, which suppresses the long-distance
contributions in the large $b$ region, and vanishes as $b=1/\Lambda$,
$\Lambda\equiv \Lambda_{\rm QCD}$ being the QCD scale. This factor
guarantees the applicability of PQCD around the energy scale of the $b$
quark mass, which is our motivation to develop a PQCD formalism for
heavy-meson decays. For a detailed derivation of the relevant Sudakov form
factors, we refer the readers to \cite{LY1,L1}.
With all the large logarithms organized,
the remaining finite contributions are absorbed into the hard $b$ quark decay
amplitude $H(t)$.

We do not claim that contributions to $B$ meson decays are purely
perturbative, but that perturbative contributions are important, and it
makes sense to compare PQCD predictions with experimental data.
It has been shown that about 70\% of the full contribution to
the decay $B\to D\pi$ arises from the region with $\alpha_s< 1$
($\alpha_s/\pi < 0.3$) in the above PQCD formalism \cite{L1}. Since energy 
releases involved in $B$ meson decays are located in the transition region 
from nonperturbative to perturbative QCD according to the analysis in 
\cite{CCL},  nonperturbative approaches, such as QCD sum rules \cite{PB}, 
also give sizeable contributions. In fact, $B$ meson decays are a subject
on which perturbative and nonperturbative approaches complement
each other.

A three-scale factorization formula for exclusive 
$B$ meson decays possesses the typical
expression,
\begin{eqnarray}
c(t)\otimes H(t)\otimes \phi(x,b)
\otimes\exp\left[-s(P,b)-2\int_{1/b}^t\frac{d{\bar\mu}}
{\bar\mu}\gamma(\alpha_s({\bar\mu}))\right],
\label{for}
\end{eqnarray}
where the exponential involving the quark anomalous dimension
$\gamma=-\alpha_s/\pi$ describes the evolution from $t$ to $1/b$
mentioned above. All the convolution factors in the above formula, except
for the wave functions $\phi(x,b)$, are calculable in perturbation theory.
The wave functions, though not calculable, are universal, since they 
absorb the long-distance dynamics, which is insensitive to
the specific decay of the $b$ quark into
light quarks with large energy release. The universality of nonperturbative
wave functions is the fundamental concept of PQCD factorization theorems.
It can be proved rigorously, but the proof is not the point of this work,
and will not be supplied here.

Because of the universality, the strategy of PQCD factorization theorems is
as follows: evaluate all  perturbative factors for some decay modes, and
adjust the wave functions such that predictions from the corresponding
factorization formulas match experimental data. At this stage, the
nonperturbative wave functions are determined up to the twists and orders of
the coupling constant, at which the factorization formulas are constructed.
Then evaluate all the perturbative factors for another decay mode. Input
the extracted wave functions into the factorization formulas of the same
twist and orders, and make predictions. With this strategy, PQCD
factorization theorems are model independent and possess a predictive power.

For light-meson wave functions, such as those of $\pi$, $\rho$, and
$K^{(*)}$ mesons, various experimental data and QCD sum rules \cite{CZ,BC,BB}
have provided reliable information. For heavy mesons, such as $B$ and
$D^{(*)}$, the knowledge is still poor. In this paper we shall determine
the heavy meson wave functions from the available experimental data. Since
two-body nonleptonic decays involve the maximal energy release that
guarantees the applicability of PQCD, and their data are abundant, the
decays $B\to D^{(*)}\pi$ are ideal processes for extracting the $B$ and
$D^{(*)}$ meson wave functions. However, both of the $B$ and $D^{(*)}$ wave
functions are unknown, and thus we need the additional data of the
$B\to K^*\gamma$ decay \cite{CLEOEX2} in order to determine the $B$ meson
wave function first. After determining the $B$ meson wave function, we
apply it to the analysis of the $B\to D^{(*)}\pi$ decays and obtain 
the $D$ and $D^{*}$ meson wave functions.

The factorization formulas for the $B\to K^*\gamma$ and $B\to D^{(*)}\pi$
decays are presented in Sec.~II and Sec.~III, respectively. The numerical
analysis is performed in Sec.~IV. Section V is the conclusion.

\section{The $B\to K^*\gamma$ Decay}

The effective Hamiltonian for the flavor-changing $b\to s$ transition is
given by \cite{REVIEW}
\begin{equation}
H_{\rm eff}=-{G_F\over
\sqrt{2}}V^*_{ts}V_{tb}\sum_{i=1}^{8}C_i(\mu)O_i(\mu),
\label{hbk}
\end{equation}
with
\begin{eqnarray}
& &O_1 = (\bar{s}_ic_j)_{V-A}(\bar{c}_jb_i)_{V-A}\;,\;\;\;\;\;\;\;\;
O_2 = (\bar{s}_ic_i)_{V-A}(\bar{c}_jb_j)_{V-A}\;, 
\nonumber \\
& &O_3 =(\bar{s}_ib_i)_{V-A}\sum_{q}(\bar{q}_jq_j)_{V-A}\;,\;\;\;\;
O_4 =(\bar{s}_ib_j)_{V-A}\sum_{q}(\bar{q}_jq_i)_{V-A}\;, 
\nonumber \\
& &O_5 =(\bar{s}_ib_i)_{V-A}\sum_{q}(\bar{q}_jq_j)_{V+A}\;,\;\;\;\;
O_6 =(\bar{s}_ib_j)_{V-A}\sum_{q}(\bar{q}_jq_i)_{V+A}\;, 
\nonumber \\
& &O_7 ={e\over
4\pi^2}\bar{s}_i\sigma^{\mu\nu}(m_sP_L+m_bP_R)b_iF_{\mu\nu}\;,
\nonumber\\
& &O_8 =-{g\over 4\pi^2}\bar{s}_i\sigma^{\mu\nu}(m_sP_L+m_bP_R)T^a_{ij}
b_jG^a_{\mu\nu},
\end{eqnarray} 
$i, \ j$ being the color indices. It has been observed that the
contributions to the $B\to K^*\gamma$ decay rate come mainly from the
operators $O_2$, $O_7$ and $O_8$ \cite{LL}. $O_2$ contributes via a 
charm-quark loop, from which the photon and an off-shell gluon are radiated. 
This off-shell gluon is then reabsorbed by the spectator quark in the $B$ or
$K^*$ meson. This $b\to sg^*\gamma$ vertex has been computed in
\cite{LL,MIL}.

The momenta of the $B$ and $K^*$ mesons in light-cone
coordinates are written as $P_1=(M_B/\sqrt{2})(1,1,{\bf 0}_T)$ and
$P_2=(M_B/\sqrt{2})(1,r^2,{\bf 0}_T)$, respectively, with $r=M_{K^*}/M_B$.
The $B$ meson is at rest with the above parametrization of momenta. We
define the momenta of light valence quarks in the $B$ and $K^*$
mesons as $k_1$ and $k_2$, respectively. $k_1$ has a minus component
$k_1^-$, giving the momentum fraction $x_1=k_1^-/P_1^-$, and small
transverse components ${\bf k}_{1T}$. $k_2$ has a large plus component
$k_2^+$, giving $x_2=k_2^+/P_2^+$, and small ${\bf k}_{2T}$. The photon
momentum is then $P_3=P_1-P_2$, whose nonvanishing component is only
$P_3^-$.

The $B\to K^*\gamma$ decay amplitude can be decomposed as
\begin{equation}
M=\epsilon^*_\gamma\cdot\epsilon^*_{K^*}M^S+
i\epsilon_{\mu\rho+-}\epsilon_\gamma^{*\mu}\epsilon^{*\rho}_{K^*}M^P\;,
\label{am}
\end{equation}
with $\epsilon_\gamma$ and $\epsilon_{K^*}$ the polarization vectors of the
photon and of the $K^*$ meson, respectively. The total rate of the
$B\to K^*\gamma$ decay is given by
\begin{equation}
\Gamma=\frac{1-r^2}{8\pi M_B}(|M^S|^2+|M^P|^2)\;.
\end{equation}
We further decompose $M^S$ and $M^P$ as
\begin{equation}
M^i=M^i_2+M^i_7+M^i_8 \;,
\end{equation}
where $i=S$ or $P$, and the terms on the right-hand side represent
the contributions from the operators $O_2$, $O_7$, and $O_8$, respectively.

The Sudakov resummation of the large logarithmic corrections to the $B$
meson wave function $\phi_B$ and to the $K^*$ meson wave function
$\phi_{K^*}$ leads to the exponentials $\exp(-S_B)$ and $\exp(-S_{K^*})$,
respectively, with the exponents
\begin{eqnarray}
S_B(t)&=&s(x_1P_1^-,b_1)+2\int_{1/b_1}^{t}\frac{d{\bar \mu}}{\bar \mu}
\gamma(\alpha_s({\bar \mu}))\;,
\nonumber \\
S_{K^*}(t)&=&s(x_2P_2^+,b_2)+s((1-x_2)P_2^+,b_2)+2\int_{1/b_2}^{t}
\frac{d{\bar \mu}}{\bar \mu}\gamma(\alpha_s({\bar \mu}))\;.
\label{sbk}
\end{eqnarray}
The variable $b_1$ ($b_2$), conjugate to the parton transverse momentum
$k_{1T}$ ($k_{2T}$), represents the transverse extent of the $B$ ($K^*$)
meson. The exponent $s$ is \cite{CS,BS}
\begin{equation}
s(Q,b)=\int_{1/b}^{Q}\frac{d \mu}{\mu}\left[\ln\left(\frac{Q}{\mu}
\right)A(\alpha_s(\mu))+B(\alpha_s(\mu))\right]\;,
\label{fsl}
\end{equation}
where the anomalous dimensions $A$ to two loops and $B$ to one loop are
\begin{eqnarray}
A&=&C_F\frac{\alpha_s}{\pi}+\left[\frac{67}{9}-\frac{\pi^2}{3}
-\frac{10}{27}n_f+\frac{2}{3}\beta_0\ln\left(\frac{e^{\gamma_E}}{2}\right)
\right]\left(\frac{\alpha_s}{\pi}\right)^2\;,
\nonumber \\
B&=&\frac{2}{3}\frac{\alpha_s}{\pi}\ln\left(\frac{e^{2\gamma_E-1}}
{2}\right)\;,
\end{eqnarray}
with $C_F=4/3$ a color factor and $\gamma_E$ the Euler constant.
The one-loop expression of the running coupling constant,
\begin{equation}
\alpha_s(\mu)=\frac{4\pi}{\beta_0\ln(\mu^2/\Lambda^2)}\;,
\end{equation}
is substituted into Eq.~(\ref{fsl}) with the coefficient
$\beta_{0}=(33-2n_{f})/3$.
The Sudakov factor in Eq.~(\ref{sbk}) suppresses long-distance contributions
from the large $b$ region, and improves the applicability of PQCD to
$B$ meson decays.

We quote the factorization formulas for $M^i_l$ \cite{LL} in terms of the
overall factor
\begin{equation}
\Gamma^{(0)}=\frac{G_F}{\sqrt{2}}\frac{e}{\pi}V^*_{ts}V_{tb}C_FM_B^5\;.
\end{equation}
The amplitudes contributed by $O_2$ are written as
\begin{eqnarray}
M^S_2&=&\Gamma^{(0)}\frac{4}{3}\int_0^1dx\int_0^{1-x}dy\int_0^1dx_1dx_2
\int_0^{1/\Lambda}bdb\phi_B(x_1,b)\phi_{K^*}(x_2)
\nonumber\\
& &\times \alpha_s(t_2)c_2(t_2)\exp[-S_B(t_2)-S_{K^*}(t_2)]_{b_1=b_2=b}
\nonumber\\
& &\times[(1-r^2+2rx_2+2x_1)y-(rx_2+3x_1)(1-x)]
\nonumber\\
& &\times\frac{(1-r)(1-r^2)x_2x}{xy(1-r^2)x_2M_B^2-M_c^2}
H_{2}(Ab,\sqrt{|B_2^2|}b)\;,
\end{eqnarray}
\begin{eqnarray}
M^P_2&=&\Gamma^{(0)}\frac{4}{3}\int_0^1dx\int_0^{1-x}dy\int_0^1dx_1dx_2
\int_0^{1/\Lambda}bdb\phi_B(x_1,b)\phi_{K^*}(x_2)
\nonumber\\
& &\times \alpha_s(t_2)c_2(t_2)\exp[-S_B(t_2)-S_{K^*}(t_2)]_{b_1=b_2=b}
\nonumber\\
& &\times \left[\left((1-r)(1-r^2)+2r^2x_2+2x_1\right)y\right.
\nonumber\\
& &\left.-\left(r(1+r)x_2+(3-r)x_1\right)(1-x)\right]
\nonumber\\
& &\times \frac{(1-r^2)x_2x}{xy(1-r^2)x_2M_B^2-M_c^2}
H_{2}(Ab,\sqrt{|B_2^2|}b)\;,
\end{eqnarray}
with
\begin{eqnarray}
A^2&=&x_1x_2M_B^2\;,
\nonumber\\
B_2^2&=&x_1x_2M_B^2-\frac{y}{1-x}(1-r^2)x_2M_B^2+\frac{M_c^2}{x(1-x)}\;,
\nonumber\\
t_2&=&\max(A,\sqrt{|B_2^2|},1/b)\;,
\end{eqnarray}
and $M_c$ being the charm quark mass.
$t_2$ is the characteristic scale of the hard amplitude
\begin{eqnarray}
H_{2}(Ab,\sqrt{|B_2^2|}b)&=&K_0(Ab)-K_0(\sqrt{|B_2^2|}b)\;,\;\;\;\;\;\;\;
B_2^2>0\;,
\nonumber\\
&=&K_0(Ab)-i\frac{\pi}{2}H_0^{(1)}(\sqrt{|B_2^2|}b)\;,\;\;\;\;
B_2^2<0\;.
\end{eqnarray}

The amplitudes from $O_7$ with a hard gluon attaching the quarks in the
$B$ meson or in the $K^*$ meson are
\begin{eqnarray}
M^S_7=-M^P_7&=&\Gamma^{(0)}2\int_0^1dx_1dx_2\int_0^{1/\Lambda}
b_1db_1b_2db_2\phi_B(x_1,b_1)\phi_{K^*}(x_2)
\nonumber\\
& &\times \alpha_s(t_7)C_7(t_7)\exp[-S_B(t_7)-S_{K^*}(t_7)]
\nonumber\\
& &\times (1-r^2)
\Bigg\{rH^{(a)}_{7}(Ab_2,\sqrt{|B_7^2|}b_1,\sqrt{|B_7^2|}b_2)
\nonumber\\
& &+[1+r+(1-2r)x_2]
H^{(b)}_{7}(Ab_1,C_7b_1,C_7b_2)\Bigg\}\;,
\end{eqnarray}
with
\begin{eqnarray}
& &B_7^2=(x_1-r^2)m_B^2\;,\;\;\;\;C_7^2=x_2m_B^2\;,
\nonumber\\
& &t_7=\max(A,\sqrt{|B_7^2|},C_7,1/b_1,1/b_2)\;.
\end{eqnarray}
The hard functions are given by
\begin{eqnarray}
& &H^{(a)}_{7}(Ab_2,\sqrt{|B_7^2|}b_1,\sqrt{|B_7^2|}b_2)=
\nonumber\\
& &\hspace{1.0cm} K_0(Ab_2)
h(\sqrt{|B_7^2|}b_1,\sqrt{|B_7^2|}b_2)\;,\;\;\;\;B_7^2>0\;,
\nonumber\\
& &\hspace{1.0cm} K_0(Ab_2)
h'(\sqrt{|B_7^2|}b_1,\sqrt{|B_7^2|}b_2)\;,\;\;\;\;B_7^2<0\;,
\end{eqnarray}
with
\begin{eqnarray}
h&=&\theta(b_1-b_2)K_0(\sqrt{|B_7^2|}b_1)
I_0(\sqrt{|B_7^2|}b_2)+(b_1\leftrightarrow b_2)\;,
\nonumber\\
h'&=&i\frac{\pi}{2}\left[\theta(b_1-b_2)
H_0^{(1)}(\sqrt{|B_7^2|}b_1)J_0(\sqrt{|B_7^2|}b_2)+
(b_1\leftrightarrow b_2)\right]\;,
\end{eqnarray}
and
\begin{eqnarray}
H^{(b)}_{7}(Ab_1,C_7b_1,C_7b_2)=K_0(Ab_1)h(C_7b_1,C_7b_2)\;.
\end{eqnarray}
The relation $M_7^S=-M_7^P$ reflects the equality of the parity-conserving
and parity-violating contributions induced by $O_7$.

The amplitudes from $O_8$, where the photon is radiated by each quark in
the $B$ and $K^*$ mesons, are
\begin{eqnarray}
M^S_8&=&-\Gamma^{(0)}\frac{1}{3}\int_0^1dx_1dx_2\int_0^{1/\Lambda}
b_1db_1b_2db_2\phi_B(x_1,b_1)\phi_{K^*}(x_2)
\nonumber\\
& &\times \alpha_s(t_8)C_8(t_8)\exp[-S_B(t_8)-S_{K^*}(t_8)]
\nonumber\\
& &\times \Bigg\{(1-r^2+x_1)(rx_2+x_1)H^{(a)}_8(Ab_2,B_8b_1,B_8b_2)
\nonumber\\
& &\hspace{0.5cm}+[(2-3r)x_2-x_1+r(1-x_2)(rx_2-2rx_1+3x_1)]
\nonumber\\
& &\hspace{1.0cm}\times H^{(b)}_8(Ab_1,C_8b_1,C_8b_2)
\nonumber\\
& &\hspace{0.5cm}+(1+r)(1-r^2)[(1+r)x_1-rx_2]
H^{(c)}_8(\sqrt{|A'^2|}b_2,D_8b_1,D_8b_2)
\nonumber\\
& &\hspace{0.5cm}
-[(1-r^2)((1-r^2)(2-x_2)+(1+3r)(2x_2-x_1))
\nonumber\\
& &\hspace{1.0cm}+2r^2x_2(x_2-x_1)]
H^{(d)}_8(\sqrt{|A'^2|}b_1,E_8b_1,E_8b_2)\Bigg\}\;,
\\
M^P_8&=&\Gamma^{(0)}\frac{1}{3}\int_0^1dx_1dx_2\int_0^{1/\Lambda}
b_1db_1b_2db_2\phi_B(x_1,b_1)\phi_{K^*}(x_2)
\nonumber\\
& &\times \alpha_s(t_8)C_8(t_8)\exp[-S_B(t_8)-S_{K^*}(t_8)]
\nonumber\\
& &\times \Bigg\{(1-r^2+x_1)(rx_2+x_1)H^{(a)}_8(Ab_2,B_8b_1,B_8b_2)
\nonumber\\
& &\hspace{0.5cm}+[(2-3r)x_2-x_1-r(1-x_2)(rx_2-2rx_1+3x_1)]
\nonumber\\
& &\hspace{1.0cm}\times
H^{(b)}_8(Ab_1,C_8b_1,C_8b_2)
\nonumber\\
& &\hspace{0.5cm}+(1-r)(1-r^2)[(1+r)x_1-rx_2]
H^{(c)}_8(\sqrt{|A'^2|}b_2,D_8b_1,D_8b_2)
\nonumber\\
& &\hspace{0.5cm}-[(1-r^2)((1-r^2)(2+x_2)-(1-3r)x_1)
\nonumber\\
& &\hspace{1.0cm} -2r^2x_2(x_2-x_1)]
H^{(d)}_8(\sqrt{|A'^2|}b_1,E_8b_1,E_8b_2)\Bigg\}\;,
\end{eqnarray}
with
\begin{eqnarray}
& &A^{\prime 2}=(1-r^2)(x_1-x_2)M_B^2\;,\;\;\;\;
B_8^2=(1-r^2+x_1)M_B^2\;,
\nonumber\\
& &C_8^2=(1-x_2)M_B^2\;,\;\;\;\;
D_8^2=(1-r^2)x_1M_B^2\;,\;\;\;\; E_8^2=(1-r^2)x_2M_B^2\;,
\nonumber\\
& &t_8=\max(A,\sqrt{|A'^2|},B_8,C_8,D_8,E_8,1/b_1,1/b_2)\;.
\end{eqnarray}
The hard functions are given by
\begin{eqnarray}
& &H^{(a)}_{8}(Ab_2,B_8b_1,B_8b_2)=
H^{(b)}_{7}(Ab_2,B_8b_1,B_8b_2)\;,
\nonumber\\
& &H^{(b)}_8(Ab_1,C_8b_1,C_8b_2)=K_0(Ab_1)h'(C_8b_1,C_8b_2)
\nonumber\\
& &H^{(c)}_8(\sqrt{|A'^2|}b_2,D_8b_1,D_8b_2)=
\nonumber\\
& &\hspace{1.0cm} K_0(\sqrt{|A'^2|}b_2)h(D_8b_1,D_8b_2)\;,
\;\;\;\;\;\;\;A'^2\ge 0\;,
\nonumber\\
& &\hspace{1.0cm} i\frac{\pi}{2}H_0^{(1)}
(\sqrt{|A'^2|}b_2)h(D_8b_1,D_8b_2)\;,\;\;\;\;A'^2< 0\;,
\nonumber\\
& &H^{(d)}_8(\sqrt{|A'^2|}b_1,E_8b_1,E_8b_2)=
\nonumber\\
& &\hspace{1.0cm} K_0(\sqrt{|A'^2|}b_1)h'(E_8b_1,E_8b_2)\;,
\;\;\;\;\;\;\;A'^2\ge 0\;,
\nonumber\\
& &\hspace{1.0cm} i\frac{\pi}{2}H_0^{(1)}
(\sqrt{|A'^2|}b_1)h'(E_8b_1,E_8b_2)\;,\;\;\;\;A'^2< 0\;.
\end{eqnarray}

Compared with the factorization formulas presented in \cite{LL}, we have
included the intrinsic (nonperturbative) $b$ dependence \cite{Kro} of the 
$B$ meson wave function $\phi_B$ . The $b$ dependence appearing in the
Sudakov factor is regarded as being perturbative. The intrinsic $b$
dependence, providing additional suppression in the large $b$ region, is
more essential for the $B$ meson, in which soft dynamics dominates and
Sudakov suppression is weaker. This dependence for the fast
recoiling $K^*$ meson is less important and neglected. In fact, the 
intrinsic $b$ dependence of the $K^*$ meson wave function is not yet
fully understood. For an analysis
of a single process such as the decay $B\to K^*\gamma$ \cite{LL}, the effect
of the intrinsic $b$ dependence is not significant, since it can always
be compensated by adjusting the $x$ dependence of $\phi_B$ under the
current accuracy of experimental data. However, this dependence turns out
to be essential, when one intends to explain data of several $B$ meson
decay modes simultaneously. This observation will become clear in
Sec.~IV.

Another modification is that we have chosen a single hard scale
$t$ for all diagrams contributing to an amplitude, which is defined
as the maximal scale appearing in these diagrams. In \cite{LL}
$t$ is chosen as the hard gluon momentum of each diagram, and thus differs
among various diagrams contributing to an amplitude. This modification
simplifies the factorization formulas for the decay $B\to K^*\gamma$.
The difference between these two schemes arises at the level of
next-to-next-to-leading
logarithms, if $B$ meson decays are indeed dominated by hard gluon exchanges.

\section{The $B\to D^{(*)}\pi$ Decays}

In this section we quote the factorization formulas for the nonleptonic
$B\to D^{(*)}\pi$ decays, whose effective Hamiltonian is given by
\begin{eqnarray}
H_{\rm eff}=\frac{G_F}{\sqrt 2}V_{ud}^{*}V_{cb}[C_1(\mu)O_1(\mu)+
C_2(\mu)O_2(\mu)]\;,
\label{eff1}
\end{eqnarray}
with the four-fermion operators
\begin{equation}
O_1=({\bar d}_iu_i)_{V-A}({\bar c}_jb_j)_{V-A}\;,\;\;\;\;
O_2=({\bar d}_iu_j)_{V-A}({\bar c}_jb_i)_{V-A}\;.
\end{equation}
Note the interchange of the definitions of
$O_1$ and $O_2$ compared with those in Eq.~(\ref{hbk}).
Similarly, we assign the momenta of the $B$ and $D^{(*)}$ mesons as
$P_1=(M_B/\sqrt{2})(1,1,{\bf 0}_T)$ and
$P_2=(M_B/\sqrt{2})(1,r^2,{\bf 0}_T)$, respectively, with
$r=M_{D^{(*)}}/M_B$, $M_{D^{(*)}}$ being the mass
of the $D^{(*)}$ meson. $\epsilon^*$ is the polarization vector of the
$D^*$ meson. The momentum fractions $x_1$ and $x_2$, and the transverse
momenta ${\bf k}_{1T}$ and ${\bf k}_{2T}$ are defined similarly as in 
Sec.~II.
The pion momentum $P_3=P_1-P_2$ is the same as the photon
momentum in Sec.~II. One of its valence quark carries the fractional
momentum $x_3P_3^-$, and small transverse momenta ${\bf k}_{3T}$.

In the present analysis  we regard the $D^{(*)}$ meson as a heavy meson. 
Hence, the Sudakov form factor $\exp(-S_{D^{(*)}})$
from the resummation of the double logarithmic corrections to the $D^{(*)}$
meson wave function $\phi_{D^{(*)}}$ has the same expression
as $\exp(-S_{B})$ but with different kinematical variables.
The Sudakov form factor $\exp(-S_\pi)$ associated with the
pion wave function $\phi_\pi$ contains the exponent \cite{LS}
\begin{eqnarray}
S_\pi(t)=s(x_3P_3^-,b_3)+s((1-x_3)P_3^-,b_3)+
2\int_{1/b_3}^{t}\frac{d{\bar \mu}}{\bar \mu}
\gamma(\alpha_s({\bar \mu}))\;.
\label{wpe}
\end{eqnarray}

The decay rates of $B\to D^{(*)}\pi$ have the expression
\begin{equation}
\Gamma_i=\frac{1}{128\pi}G_F^2|V_{cb}|^2|V_{ud}|^2M_B^3\frac{(1-r^2)^3}{r}
|{\cal M}_i|^2\;,
\label{dr}
\end{equation}
where $i=1$, 2, 3, and 4 denote the modes $B^-\to D^0\pi^-$,
${\bar B}^0\to D^+\pi^-$, $B^-\to D^{*0}\pi^-$, and
${\bar B}^0\to D^{*+}\pi^-$, respectively. The decay
amplitudes ${\cal M}_i$ are written as
\begin{eqnarray}
{\cal M}_1&=&f_\pi[(1+r)\xi_+-(1-r)\xi_-]+
f_D\xi_{\rm int}+{\cal M}_{\rm ext}+{\cal M}_{\rm int}\;,
\label{M1}\\
{\cal M}_2&=&f_\pi[(1+r)\xi_+-(1-r)\xi_-]+f_B
\xi_{\rm exc}+{\cal M}_{\rm ext}+{\cal M}_{\rm exc}\;,
\label{M2}\\
{\cal M}_3&=&\frac{1+r}{2r}f_\pi[(1+r)\xi_{A_1}-(1-r)(r\xi_{A_2}
+\xi_{A_3})]
\nonumber \\
& &+f_{D^*}\xi^*_{\rm int}+{\cal M}^*_{\rm ext}+{\cal M}^{*}_{\rm int}\;,
\label{M3}\\
{\cal M}_4&=&\frac{1+r}{2r}f_\pi[(1+r)\xi_{A_1}-(1-r)(r\xi_{A_2}
+\xi_{A_3})]
\nonumber \\
& &+f_B\xi^*_{\rm exc}+{\cal M}^*_{\rm ext}+{\cal M}^{*}_{\rm exc}\;,
\label{M4}
\end{eqnarray}
where $f_B$, $f_{D^{(*)}}$, and $f_\pi$ are the $B$ meson, $D^{(*)}$ meson,
and pion decay constants, respectively. The form factors $\xi_i$, $i=+$,
$-$, $V$, $A_1$, $A_2$, and $A_3$, denote the factorizable external
$W$-emission contributions. The form factors $\xi^{(*)}_{\rm int}$ and
$\xi^{(*)}_{\rm exc}$ denote the factorizable internal $W$-emission and
$W$-exchange contributions, respectively.
The amplitudes ${\cal M}_{\rm ext}^{(*)}$,
${\cal M}_{\rm int}^{(*)}$, and ${\cal M}_{\rm exc}^{(*)}$ represent the
nonfactorizable external $W$-emission, internal $W$-emission, and
$W$-exchange contributions, respectively.

The form factors $\xi_i$, $i=+$, $A_1$ and $A_3$, and $\xi_j$, $j=-$ and
$A_2$, are given by
\begin{eqnarray}
\xi_i&=& 16\pi C_F\sqrt{r}M_B^2
\int_{0}^{1}d x_{1}d x_{2}\int_{0}^{\infty} b_1d b_1 b_2d b_2
\phi_B(x_1,b_1)\phi_{D^{(*)}}(x_2,b_2)
\nonumber \\
& &\times\alpha_s(t)a_1(t)\exp[-S_B(t)-S_{D^{(*)}}(t)]
\nonumber \\
& &\times [(1+\zeta_ix_2r)h(x_1,x_2,b_1,b_2,m)
+(r+\zeta'_ix_1)h(x_2,x_1,b_2,b_1)]\;,
\label{+}\\
\xi_j&=& 16\pi C_F\sqrt{r}M_B^2
\int_{0}^{1}d x_{1}d x_{2}\int_{0}^{\infty} b_1d b_1 b_2d b_2
\phi_B(x_1,b_1)\phi_{D^{(*)}}(x_2,b_2)
\nonumber \\
& &\times \alpha_s(t)a_1(t)\exp[-S_B(t)-S_{D^{(*)}}(t)]
\nonumber \\
& &\times [\zeta_j x_2rh(x_1,x_2,b_1,b_2)
+\zeta'_jx_1 h(x_2,x_1,b_2,b_1)]\;,
\label{-}
\end{eqnarray}
with the constants \cite{WYL1}
\begin{eqnarray}
& &\zeta_+=\zeta'_+=\frac{1}{2}\left[\eta-\frac{3}{2}+
\sqrt{\frac{\eta-1}{\eta+1}}\left(\eta-\frac{1}{2}\right)\right]\;,
\nonumber \\
& &\zeta_-=-\zeta'_-=-\frac{1}{2}\left[\eta-\frac{1}{2}+
\sqrt{\frac{\eta+1}{\eta-1}}\left(\eta-\frac{3}{2}\right)\right]\;,
\nonumber \\
& &\zeta_{A_1}=-\frac{2-\eta-\sqrt{\eta^2-1}}{2(\eta+1)}\;,
\;\;\;\;\zeta'_{A_1}=-\frac{1}{2(\eta+1)}\;,
\nonumber \\
& &\zeta_{A_2}=0\;,\;\;\;\;\zeta'_{A_2}=-1-\frac{\eta}{\sqrt{\eta^2-1}}\;,
\nonumber \\
& &\zeta_{A_3}=-\frac{1}{2}-\frac{\eta-2}{2\sqrt{\eta^2-1}}\;,
\;\;\;\;\zeta'_{A_3}=\frac{1}{2\sqrt{\eta^2-1}}\;,
\end{eqnarray}
where the velocity transfer $\eta$ takes the maximal value
$\eta_{\max}=(1+r^2)/(2r)$ here.
The form factors $\xi^{(*)}_{\rm int}$ and $\xi^{(*)}_{\rm exc}$
are written as
\begin{eqnarray}
\xi^{(*)}_{\rm int}&=&16\pi C_F\sqrt{r}M_B^2
\int_0^1 dx_1dx_3\int_0^{\infty}b_1db_1b_3db_3
\phi_B(x_1,b_1)\phi_\pi(x_3)
\nonumber \\
& &\times \alpha_s(t_{\rm int})a_2(t_{\rm int})
\exp[-S_B(t_{\rm int})-S_\pi(t_{\rm int})]
\nonumber \\
& &\times\left[(1+x_3(1-r^2))h_{\rm int}(x_1,x_3,b_1,b_3,m_{\rm int})
\right.
\nonumber \\
& &\left.+\zeta^{(*)}_{\rm int}x_1r^2
h_{\rm int}(x_3,x_1,b_3,b_1,m_{\rm int})\right]\;,
\label{int} \\\
\xi^{(*)}_{\rm exc}&=&16\pi C_F\sqrt{r}M_B^2
\int_0^1 dx_2dx_3\int_0^{\infty}b_2db_2b_3db_3
\phi_{D^{(*)}}(x_2,b_2)\phi_\pi(x_3)
\nonumber \\
& &\times\alpha_s(t_{\rm exc})a_2(t_{\rm exc})
\exp[-S_{D^{(*)}}(t_{\rm exc})-S_\pi(t_{\rm exc})]
\nonumber \\
& &\times
\left[(x_3(1-r^2)-\zeta^{(*)}_{\rm exc}r^2)
h_{\rm exc}(x_2,x_3,b_2,b_3,m_{\rm exc})\right.
\nonumber \\
& &\left.-x_2h_{\rm exc}(x_3,x_2,b_3,b_2,m_{\rm exc})\right]\;,
\label{exc}
\end{eqnarray}
with the constants $\zeta_{\rm int}=\zeta_{\rm exc}=-\zeta^*_{\rm int}
=-\zeta^*_{\rm exc}=1$.

In Eqs.~(\ref{+}), (\ref{-}), (\ref{int}), and (\ref{exc}) the Wilson
coefficients $a_1$ and $a_2$ are defined by
\begin{equation}
a_1=C_1+\frac{C_2}{N_c}\;,\;\;\;\;
a_2=C_2+\frac{C_1}{N_c}\;,
\end{equation}
with $N_c=3$ the number of colors. The functions $h$'s, obtained from the 
Fourier transformation of the lowest-order $H$, are given by
\begin{eqnarray}
h(x_1,x_2,b_1,b_2,m)&=&K_{0}\left(\sqrt{x_1x_2m}b_1\right)
\nonumber \\
& &\times \left[\theta(b_1-b_2)K_0\left(\sqrt{x_2m}
b_1\right)I_0\left(\sqrt{x_2m}b_2\right)\right.
\nonumber \\
& &\left.+\theta(b_2-b_1)K_0\left(\sqrt{x_2m}b_2\right)
I_0\left(\sqrt{x_2m}b_1\right)\right],
\label{dh}\\
h_{\rm int}(x_1,x_3,b_1,b_3,m_{\rm int})&=&h(x_1,x_3,b_1,b_3,
m_{\rm int})\;,
\label{hint}\\
h_{\rm exc}(x_2,x_3,b_2,b_3,m_{\rm exc})&=&\frac{\pi^2}{4}
H_0^{(1)}\left(\sqrt{x_2x_3m_{\rm exc}}b_2\right)
\nonumber \\
& &\times\left[\theta(b_2-b_3)
H_0^{(1)}\left(\sqrt{x_3m_{\rm exc}}b_2\right)
J_0\left(\sqrt{x_3m_{\rm exc}}b_3\right)\right.
\nonumber \\
& &\left.+\theta(b_3-b_2)H_0^{(1)}\left(\sqrt{x_3m_{\rm exc}}b_3\right)
J_0\left(\sqrt{x_3m_{\rm exc}}b_2\right)\right],
\nonumber \\
& &
\end{eqnarray}
with $m=M_B^2$ and $m_{\rm int}=m_{\rm exc}=(1-r^2)M_B^2$.
The hard scales $t$ are chosen as
\begin{eqnarray}
t&=&{\rm max}(\sqrt{x_1m},\sqrt{x_2m},1/b_1,1/b_2)\;, \\
t_{\rm int}&=&{\rm max}(\sqrt{x_1m_{\rm int}},\sqrt{x_3m_{\rm int}},
1/b_1,1/b_3)\;, \\
t_{\rm exc}&=&{\rm max}(\sqrt{x_2m_{\rm exc}},\sqrt{x_3m_{\rm exc}},
1/b_2,1/b_3)\;.
\end{eqnarray}

For the nonfactorizable amplitudes, the factorization formulas involve
the kinematic variables of all the three mesons, and the Sudakov exponent
is given by $S=S_B+S_{D^{(*)}}+S_\pi$. The integration over $b_3$ can be
performed trivially, leading to $b_3=b_1$ or $b_3=b_2$. Their
expressions are
\begin{eqnarray}
{\cal M}^{(*)}_{\rm ext}&=& 32\pi\sqrt{2N} C_F\sqrt{r}M_B^2
\int_0^1 [dx]\int_0^{\infty}
b_1 db_1 b_2 db_2
\phi_B(x_1,b_1)\phi_{D^{(*)}}(x_2,b_2)\phi_\pi(x_3)
\nonumber \\
& &\times \alpha_s(t_b)\frac{C_2(t_b)}{N}\exp[-S(t_b)|_{b_2=b_1,b_3=b_2}]
\nonumber \\
& &\times \biggl\{
[x_3(1-r^2)-x_1-\zeta^{(*)}_bx_2(r-r^2)]h^{(1)}_b(x_i,b_i)
\nonumber \\
& &\hspace{0.5cm}
-[x_3(1-r^2)-x_1+x_2]h^{(2)}_b(x_i,b_i) \biggr\}\;,
\label{mb}\\
{\cal M}^{(*)}_{\rm int}&=& 32\pi\sqrt{2N} C_F\sqrt{r}M_B^2
\int_0^1 [dx]\int_0^{\infty}b_1 db_1 b_2 db_2
\phi_B(x_1,b_1)\phi_{D^{(*)}}(x_2,b_2)\phi_\pi(x_3)
\nonumber \\
& &\times \alpha_s(t_d)\frac{C_1(t_d)}{N}\exp[-S(t_d)|_{b_3=b_1}]
\nonumber \\
& &\times 
\biggl\{[x_1-x_2-x_3(1-r^2)]h^{(1)}_d(x_i,b_i)
\nonumber \\
& &\hspace{0.5cm}
-[(x_1+x_2)(1+\zeta^{(*)}_d r^2)-1]h^{(2)}_d(x_i,b_i)
\biggr\}\;,
\label{md}\\
{\cal M}^{(*)}_{\rm exc}&=& 32 \pi\sqrt{2N} C_F\sqrt{r}M_B^2
\int_0^1 [dx]\int_0^{\infty}b_1 db_1 b_2 db_2
\phi_B(x_1,b_1)\phi_{D^{(*)}}(x_2,b_2)\phi_\pi(x_3)
\nonumber \\
& &\times \alpha_s(t_f)\frac{C_1(t_f)}{N}\exp[-S(t_f)|_{b_3=b_2}]
\nonumber \\
& &\times \biggl\{
[x_3(1-r^2)-\zeta^{(*)}_f(x_1-x_2)r^2]h^{(1)}_f(x_i,b_i)
\nonumber \\
& &\hspace{0.5cm}
-[(x_1+x_2)(1+\zeta^{(*)}_fr^2)
-\zeta^{(*)}_fr^2]h^{(2)}_f(x_i,b_i) \biggr\}\;,
\label{mf}
\end{eqnarray}
with the definition $[dx]\equiv dx_1dx_2dx_3$.
The constants are $\zeta_{b,d,f}=-\zeta^{*}_{b,d,f}=1$.

The functions $h^{(j)}$, $j=1$ and 2, appearing in
Eqs.~(\ref{mb})-(\ref{mf}), are written as
\begin{eqnarray}
\everymath{\displaystyle}
h^{(j)}_b&=& \left[\theta(b_1-b_2)K_0\left(BM_B
b_1\right)I_0\left(BM_Bb_2\right)\right. \nonumber \\
& &\quad \left.
+\theta(b_2-b_1)K_0\left(BM_B b_2\right)
I_0\left(BM_B b_1\right)\right]\;  \nonumber \\
&  & \times \left( \begin{array}{cc}
 K_{0}(B_{j}M_Bb_{2}) &  \mbox{for $B^2_{j} \geq 0$}  \\
 \frac{i\pi}{2} H_{0}^{(1)}(\sqrt{|B_{j}^2|}M_Bb_{2})  &
 \mbox{for $B^2_{j} \leq 0$}
  \end{array} \right)\;,          
\\
\everymath{\displaystyle}
h^{(j)}_d&=& \left[\theta(b_1-b_2)K_0\left(DM_B
b_1\right)I_0\left(DM_Bb_2\right)\right. \nonumber \\
& &\quad \left.
+\theta(b_2-b_1)K_0\left(DM_B b_2\right)
I_0\left(DM_B b_1\right)\right]\;  \nonumber \\
&  & \times \left( \begin{array}{cc}
 K_{0}(D_{j}M_Bb_{2}) &  \mbox{for $D^2_{j} \geq 0$}  \\
 \frac{i\pi}{2} H_{0}^{(1)}(\sqrt{|D_{j}^2|}M_Bb_{2})  &
 \mbox{for $D^2_{j} \leq 0$}
  \end{array} \right)\;,          
\label{hjd}\\
\everymath{\displaystyle}
h^{(j)}_f&=& i\frac{\pi}{2}
\left[\theta(b_1-b_2)H_0^{(1)}\left(FM_B
b_1\right)J_0\left(FM_Bb_2\right)\right. \nonumber \\
& &\quad\left.
+\theta(b_2-b_1)H_0^{(1)}\left(FM_B b_2\right)
J_0\left(FM_B b_1\right)\right]\;  \nonumber \\
&  & \times \left( \begin{array}{cc}
 K_{0}(F_{j}M_Bb_{1}) &  \mbox{for $F^2_{j} \geq 0$}  \\
 \frac{i\pi}{2} H_{0}^{(1)}(\sqrt{|F_{j}^2|}M_Bb_{1})  &
 \mbox{for $F^2_{j} \leq 0$}
  \end{array} \right)\;,          
\end{eqnarray}
with the variables
\begin{eqnarray}
B^{2}&=&x_{1}x_{2}\;,
\nonumber \\
B_{1}^{2}&=&x_{1}x_{2}-x_2x_{3}(1-r^{2})\;,
\nonumber \\
B_{2}^{2}&=&x_{1}x_{2}(1+r^{2})-(x_{1}-x_{2})(1-x_{3})(1-r^{2})\;,
\nonumber \\
D^{2}&=&x_{1}x_{3}(1-r^{2})\;,
\nonumber \\
D_{1}^{2}&=&F_1^2=3D(x_{1}-x_{2})x_{3}(1-r^{2})\;,
\nonumber \\
D_{2}^{2}&=&(x_{1}+x_{2})r^{2}-(1-x_{1}-x_{2})x_{3}(1-r^{2})\;,
\nonumber \\
F^{2}&=&x_{2}x_{3}(1-r^{2})\;,
\nonumber \\
F_{2}^{2}&=&x_{1}+x_{2}+(1-x_{1}-x_{2})x_{3}(1-r^{2})\;.
\label{mis}
\end{eqnarray}
The scales $t^{(j)}$ are chosen as
\begin{eqnarray}
t_b&=&{\rm max}(BM_B,\sqrt{|B_1^2|}M_B,\sqrt{|B_2^2|}M_B,
1/b_1,1/b_2)\;,
\nonumber \\
t_d&=&{\rm max}(DM_B,\sqrt{|D_1^2|}M_B,\sqrt{|D_2^2|}M_B,
1/b_1,1/b_2)\;,
\nonumber \\
t_f&=&{\rm max}(FM_B,\sqrt{|F_1^2|}M_B,\sqrt{|F_2^2|}M_B,
1/b_1,1/b_2)\;.
\end{eqnarray}

As before, we have included the intrinsic $b$ dependence of wave functions 
for the heavy $B$ and
$D^{(*)}$ mesons, in which soft dynamics is more important. This dependence 
for the fast recoiling pion is also neglected, similar to the case of 
the $K^*$ meson, for consistency. The scheme of
a single hard scale $t$ for all diagrams contributing to an
amplitude has been adopted to simplify the factorization formulas. We have
also corrected misprints in Eq.~(\ref{mis}) appearing in \cite{YL}.

\section{Numerical Analysis}

In this section we determine the heavy-meson wave functions by fitting
the predictions from the factorization formulas for the
$B\to K^*\gamma$ and $B\to D^{(*)}\pi$ decays to the CLEO data
\cite{CLEOEX2,TEB,GB}
\begin{eqnarray}
B(B\to K^*\gamma)&=&(4.2\pm 0.8\pm 0.6)\times 10^{-5}\;,
\nonumber\\
B(B^-\to D^0\pi^-)&=&(5.0\pm 0.5\pm 0.2)\times 10^{-3}\;,
\nonumber\\
B({\bar B}^0\to D^+\pi^-)&=&(3.1\pm 0.4\pm 0.2)\times 10^{-3}\;,
\nonumber\\
B(B^-\to D^{*0}\pi^-)&=&(4.34\pm 0.33\pm 0.34\pm 0.18)\times 10^{-3}\;,
\nonumber\\
B({\bar B}^0\to D^{*+}\pi^-)&=&(2.81\pm 0.11\pm 0.21\pm 0.05)\times
10^{-3}\;.
\label{dat}
\end{eqnarray}
We adopt $G_F=1.16639\times 10^{-5}$ GeV$^{-2}$, the
CKM matrix elements $|V_{cb}|=0.040$ \cite{L1,WYL1} and $|V_{ud}|=0.974$,
the masses $M_B=5.28$ GeV, $M_D=1.87$ GeV, and $M_{D^*}=2.01$ GeV
\cite{PDG}, and the ${\bar B}^0$ ($B^-$) meson lifetime $\tau_{B^0}=1.53$
($\tau_{B^-}=1.68$) ps \cite{B}.

As the transverse extent $b$ approaches zero, $\phi(x,b)$ reduces to the
standard parton model $\phi(x)$, {\it i.e.}, $\phi(x)=\phi(x,b=0)$.
The wave functions $\phi_i(x)$, $i=B$, $D$, $D^*$, $K^*$, and $\pi$,
satisfy the normalization
\begin{equation}
\int_0^1\phi_i(x)dx=\frac{f_i}{2\sqrt{6}}\;,
\label{dco}
\end{equation}
with the corresponding decay constants $f_i$. We choose $f_B=180$ MeV
(in the convention $f_\pi=132$ MeV), which is reasonable according to the
lattice calculations presented in the literature \cite{AK}.
For the wave function of the $K^*$ meson with transverse polarizations,
we employ \cite{BB}
\begin{equation}
\phi_{K^*}(x)=\frac{\sqrt{6}}{2}f_{K^*}
x(1-x)[1+0.6(1-2x)+0.06(5(1-2x)^2-1)]\;,
\label{mn}
\end{equation}
with $f_{K^*}=185$ MeV. As to the pion wave function, we employ the
asymptotic form,
\begin{eqnarray}
\phi_\pi(x)=\frac{\sqrt{6}}{2}f_\pi x(1-x)\;,
\label{pwf}
\end{eqnarray}
which has been found to give predictions consistent with the data of
the $\pi\gamma$ form factor \cite{KM}.

For the $B$ meson wave functions, we use the model \cite{BW}
\begin{eqnarray}
\phi_B(x,b)=N_B\sqrt{x(1-x)}
\exp\left[-\frac{1}{2}\left(\frac{xM_B}{\omega_B}\right)^2
-\frac{\omega_B^2 b^2}{2}\right]\;,
\label{os}
\end{eqnarray}
where $N_B$ is the normalization constant and $\omega_B$ the shape
parameter. Another model with a flat profile \cite{YS}
\begin{eqnarray}
\phi'_B(x,b)=\frac{N'_Bbx^2(1-x)^2}{\sqrt{M_B^2x+C_Bx(1-x)}}
K_1\left(\sqrt{M_B^2x+C_Bx(1-x)}b\right)\;,
\label{bw}
\end{eqnarray}
gives predictions for the branching ratio of the $B\to K^*\gamma$ decay,
which are too small compared with the data. Therefore, we shall not
consider it in the numerical analysis below.
For the $D^{(*)}$ meson wave functions, we adopt the same model
as of the $B$ meson in Eq.~(\ref{os}), but with different
normalization constants $N_{D^{(*)}}$ and shape parameters
$\omega_{D^{(*)}}$,
\begin{eqnarray}
\phi_{D^{(*)}}(x,b)=N_{D^{(*)}}\sqrt{x(1-x)}
\exp\left[-\frac{1}{2}\left(\frac{xM_{D^{(*)}}}{\omega_{D^{(*)}}}\right)^2
-\frac{\omega_{D^{(*)}}^2 b^2}{2}\right]\;.
\label{osd}
\end{eqnarray}

We shall determine 5 parameters $N_D(f_D)$, $N_{D^*}(f_{D^*})$, $\omega_B$,
$\omega_D$, and $\omega_{D^*}$ from the data of the 5 processes in
Eq.~(\ref{dat}), under the constraints $f_{D^{(*)}}>f_B$, which is
reasonable from the viewpoint of heavy quark symmetry. The fitting
procedures are as follows. Start with the $K^*$ meson wave function
in Eq.~(\ref{mn}), and determine the $B$ meson
wave function ({\it i.e.}, the shape parameter $\omega_B$) from the data of
the decay $B\to K^*\gamma$. The obtained $B$ meson wave function is then
employed to evaluate the branching ratios of the charged and neutral $B$
decays $B\to D^{(*)}\pi$, from which both parameters, $N_{D^{(*)}}$
and $\omega_{D^{(*)}}$, of the $D^{(*)}$ meson wave function can be
extracted. We have tested the heavy meson wave functions
without the intrinsic $b$ dependence. It turns out that the $D^{(*)}$ meson
wave function $\phi_{D^{(*)}}(x)$, along with the $B$ meson wave function
$\phi_B(x)$ determined from the $B\to K^*\gamma$ data, always lead to
predictions for the $B\to D^{(*)}\pi$ decays which exceed the data by a
factor 3, no matter how $\phi_{D^{(*)}}(x)$ is adjusted under the constraint
$f_{D^{(*)}}>f_B$. These predictions are lowered only by introducing the
intrinsic $b$ dependence into heavy meson wave functions, implying that the
intrinsic $b$ dependence is indeed essential.

It is easy to understand that the parameters $\omega_{D^{(*)}}$ control the
difference between the charged and neutral $B$ meson decay rates. This fact
is obvious from Eqs.~(\ref{M1})-(\ref{M4}): by varying $\omega_{D^{(*)}}$,
the external $W$-emission contributions, {\it i.e.}, the $B\to D$ form
factors change, while the internal $W$-emission contributions, {\it i.e.},
the $B\to \pi$ form factors remain the same. The $W$-exchange and
nonfactorizable diagrams give smaller contributions than the $W$-emission
diagrams to the decays considered. Therefore, the charged $B$
decay rates, as a combination of the external and internal $W$-emission
contributions, and the neutral $B$-meson decay rates with the external
$W$ emissions as the main contribution, vary differently with the change of
$\omega_{D^{(*)}}$. 

\begin{figure}
\centerline{\epsfig{file=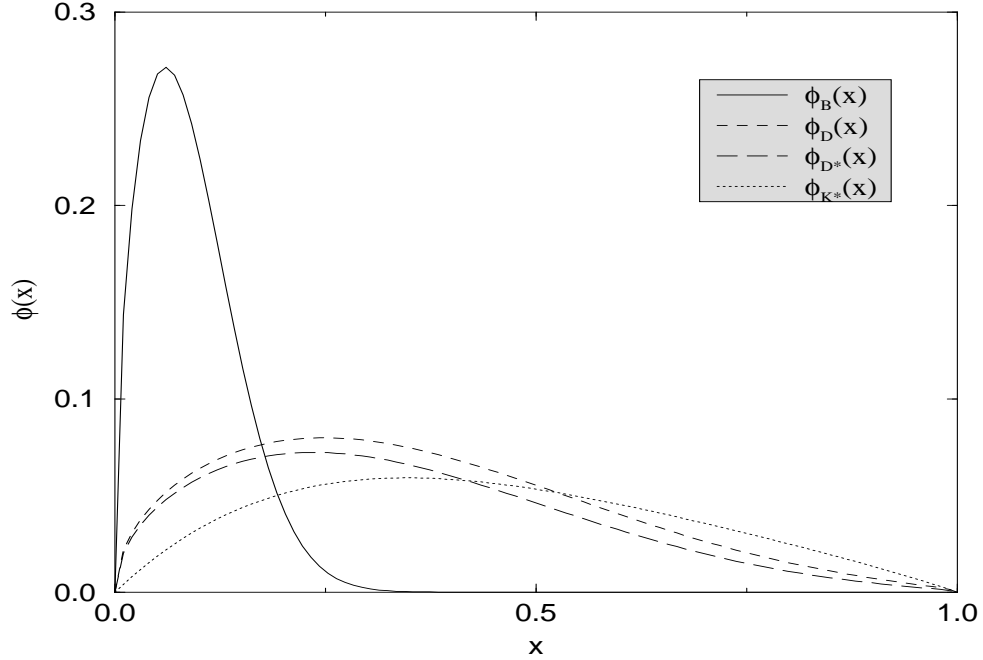,height=9cm,width=13cm,silent=3.}}
\caption{The $K^*$, $B$ and $D^{(*)}$ meson wave functions.}
\label{f:fig1}
\end{figure}

After a tedious numerical analysis, we determine the meson wave functions,
\begin{eqnarray}
\phi_B(x,b)&=&1.44605\sqrt{x(1-x)}
\exp\left[-\frac{1}{2}\left(\frac{xM_B}{0.462\;\;{\rm GeV}}\right)^2
-\frac{1}{2}(0.462\;\;{\rm GeV})^2 b^2\right]\;,
\nonumber\\
\phi_D(x,b)&=&0.218677\sqrt{x(1-x)}
\exp\left[-\frac{1}{2}\left(\frac{xM_D}{0.8\;\;{\rm GeV}}\right)^2
-\frac{1}{2}(0.8\;\;{\rm GeV})^2 b^2\right]\;,
\nonumber\\
\phi_{D^*}(x,b)&=&0.202997\sqrt{x(1-x)}
\exp\left[-\frac{1}{2}\left(\frac{xM_{D^*}}{0.8\;\;{\rm GeV}}\right)^2
-\frac{1}{2}(0.8\;\;{\rm GeV})^2 b^2\right]\;,
\label{hmw}
\end{eqnarray}
which correspond to the decay constants $f_B=180$ MeV, $f_D= 220$
MeV, and $f_{D^*}=190$ MeV, respectively. With these heavy meson wave
functions, we obtain
\begin{eqnarray}
B(B\to K^*\gamma)&=&4.19\times 10^{-5}\;,
\nonumber\\
B(B^-\to D^0\pi^-)&=&4.74\times 10^{-3}\;,
\nonumber\\
B({\bar B}^0\to D^+\pi^-)&=&3.50\times 10^{-3}\;,
\nonumber\\
B(B^-\to D^{*0}\pi^-)&=&3.93\times 10^{-3}\;,
\nonumber\\
B({\bar B}^0\to D^{*+}\pi^-)&=&2.94\times 10^{-3}\;,
\label{dat1}
\end{eqnarray}
which are consistent with the data in Eq.~(\ref{dat}).
Note that the value
$\omega_B=0.462$ GeV is in good agreement with $\omega_B=0.4$ GeV
obtained in the oscillating quark model \cite{BW}. It indicates
that meson wave functions in PQCD factorization theorems have absorbed
nonperturbative dynamics in $B$ meson decays appropriately, and that 
remained dynamics, collected by the hard quark-level amplitudes, is 
perturbative. The $x$ dependence of the extracted heavy meson wave 
functions and of the $K^*$ meson wave function is displayed in 
Fig.~1. As expected, the maximum of
a heavy meson wave function moves to the large $x$ region as the meson
mass decreases.

\section{Conclusion}

In this paper we have extracted heavy meson wave functions, as shown
in Eq.~(\ref{hmw}), from the experimental data of the
$B\to K^*\gamma$ and $B\to D^{(*)}\pi$ decays. The extracted meson
decay constants $f_D$ and $f_{D^*}$, compared to $f_B$, and the behavior 
of the wave functions exhibited in Fig.~1 are all reasonable. We have also
confirmed that the intrinsic $b$ dependence of heavy meson wave functions,
which makes possible a simultaneous fit to the data of several decay modes,
is essential. Although the uncertainties of the data are still large, such
that wide ranges of the relevant parameters are allowed, we shall employ the
obtained wave functions to make predictions for other nonleptonic $B$ meson
decay modes, especially charmless decays. As the precision of data improves,
the heavy meson wave functions determined in our formalism will be more
reliable. On the other hand, we can also proceed with a global determination
of the heavy meson wave functions by incorporating more decay modes into the
fitting procedures. These subjects will be discussed elsewhere.

\vskip 1.0cm
\acknowledgments

We thank P. Ball for useful comments and discussions.
This work was supported by the National Science Council of R.O.C.
under the Grant No. NSC-88-2112-M-006-013 and by the Ministry of Science
and Technology of the Republic of Croatia under Contract No. 00980102.


\end{document}